\documentstyle [12pt] {article}

\catcode`@=12
\begin{document}
\thispagestyle{empty}
\begin{flushright} UCRHEP-T266\\November 1999\
\end{flushright}
\vskip 0.5in
\begin{center}
{\Large \bf Leptogenesis from R Parity Nonconservation\\}
\vskip 2.0in
{\bf Thomas Hambye$^1$, Ernest Ma$^2$, Utpal Sarkar$^3$\\}
\vskip 0.3in
{$^1$ \sl INFN, Laboratori Nazionali di Frascati, I-00044 Frascati, Italy\\}
\vskip 0.1in
{$^2$ \sl Department of Physics, Univeristy of California, Riverside, 
CA 92521, USA\\}
\vskip 0.1in
{$^3$ \sl Physical Research Laboratory, Ahmedabad 380 009, India\\}
\end{center}
\vskip 1.8in
\begin{abstract}\
It is known that realistic neutrino masses for neutrino oscillations may be 
obtained from R parity nonconserving supersymmetry.  It is also known that 
such interactions would erase any preexisting lepton or baryon asymmetry of 
the Universe because of the inevitable intervention of the electroweak 
sphalerons. We now show how a crucial subset of these 
R parity nonconserving terms may in fact create its own successful 
leptogenesis.
\end{abstract}

\newpage
\baselineskip 24pt

To obtain naturally small Majorana neutrino masses\cite{ma}, the canonical 
approach is to have heavy singlet right-handed neutrinos in the guise of the 
famous seesaw mechanism\cite{seesaw}.  An equally simple alternative is to use 
a heavy scalar Higgs triplet\cite{triplet}.  Both offer the important 
additional benefit of generating a lepton asymmetry\cite{triplet,fy} which 
gets converted into the present observed baryon asymmetry of the Universe by 
virtue of the electroweak sphalerons\cite{krs,bary}.

In the past several years, a third way has come under active consideration, 
namely that of R parity nonconserving supersymmetry.  Whereas realistic 
neutrino masses for neutrino oscillations may be generated\cite{numass}, 
the lepton-number nonconserving interactions involved would certainly erase 
any preexisting lepton asymmetry of the Universe\cite{erase}.  This negative 
aspect of R parity nonconserving supersymmetry has prevented it from being 
considered as a truly competitive alternative to the canonical seesaw or 
the triplet Higgs mechanism for neutrino masses.  To remedy this situation, 
we now show for the first time how a crucial subset of these R parity 
nonconserving terms may in fact create its own successful leptogenesis, 
through the suppressed decay of the lightest neutralino into a charged 
Higgs boson and a lepton.

In a supersymmetric extension of the minimal standard model of fundamental 
interactions, the R parity of a particle is conventionally defined as
\begin{equation}
{\rm R} \equiv (-1)^{3B+L+2J},
\end{equation}
where $B$ is its baryon number, $L$ its lepton number, and $J$ its spin 
angular momentum.  Hence the standard-model particles have R = +1 and 
their supersymmetric partners have R = $-1$.  Using the common notation 
where all chiral superfields are considered left-handed, the three 
families of leptons and quarks are given by
\begin{eqnarray}
&& L_i = (\nu_i, e_i) \sim (1,2,-1/2), ~~~ e^c_i \sim (1,1,1), \\
&& Q_i = (u_i, d_i) \sim (3,2,1/6), \\
&& u^c_i \sim (3^*,1,-2/3), ~~~ d^c_i \sim (3^*,1,1/3),
\end{eqnarray}
where $i$ is the family index, and the two Higgs doublets are given by
\begin{eqnarray}
&& H_1 = (h_1^0,h_1^-) \sim (1,2,-1/2), \\
&& H_2 = (h_2^+,h_2^0) \sim (1,2,1/2),
\end{eqnarray}
where the $SU(3)_C \times SU(2)_L \times U(1)_Y$ content of each superfield 
is also indicated.  If R parity is conserved, the superpotential is 
restricted to have only the terms
\begin{eqnarray}
W &=& \mu H_1 H_2 + f^e_{ij} H_1 L_i e^c_j + f^d_{ij} H_1 Q_i d^c_j 
\nonumber \\ && + f^u_{ij} H_2 Q_i u^c_j.
\end{eqnarray}
If R parity is violated but not baryon number, then the superpotential 
contains the additional terms
\begin{equation}
W' = \mu_i L_i H_2 + \lambda_{ijk} L_i L_j e^c_k + \lambda'_{ijk} 
L_i Q_j d^c_k,
\end{equation}
resulting in nonzero neutrino masses either from mixing with the neutralino 
mass matrix or in one-loop order\cite{numass}.

Consider the simplified case\cite{maroy,cf} of having only $L_\tau$ and 
$\tau^c$ as effective $L=0$ superfields.  The neutralino mass matrix now 
spans five fields: the U(1) gaugino ($\tilde B$), the SU(2) gaugino 
($\tilde W_3$), the two Higgsinos ($\tilde h_1^0$, $\tilde h_2^0$), and 
$\nu_\tau$.
\begin{equation}
{\cal M} = \left[ \begin{array} {c@{\quad}c@{\quad}c@{\quad}c@{\quad}c} 
M_1 & 0 & -sm_3 & sm_4 & -sm_5 \\ 0 & M_2 & cm_3 & -cm_4 & cm_5 \\ 
-sm_3 & cm_3 & 0 & -\mu & 0 \\ sm_4 & -cm_4 & -\mu & 0 & -\mu_\tau \\ 
-sm_5 & cm_5 & 0 & -\mu_\tau & 0 \end{array} \right],
\end{equation}
where $s = \sin \theta_W$, $c = \cos \theta_W$, $m_3 = M_Z \cos \beta \cos 
\theta_\tau$, $m_4 = M_Z \sin \beta$, $m_5 = M_Z \cos \beta \sin \theta_\tau$, 
with $\tan \beta = \langle h_2^0 \rangle / [\langle h_1^0 \rangle^2 + 
\langle \tilde \nu_\tau \rangle^2]^{1/2}$ and $\tan \theta_\tau = \langle 
\tilde \nu_\tau \rangle / \langle h_1^0 \rangle$.

To understand the structure of the above $5 \times 5$ mass matrix, let us 
assume that $\mu$ is the dominant term, then $\tilde h^0_{1,2}$ form 
a heavy Dirac particle of mass $\mu$ and the reduced $3 \times 3$ matrix 
in the basis ($\tilde B, \tilde W_3, \nu_\tau$) becomes
\begin{equation}
{\cal M} = \left[ \begin{array} {c@{\quad}c@{\quad}c} M_1 - s^2 \delta & 
sc \delta & -s \epsilon \\ sc \delta & M_2 - c^2 \delta & c \epsilon \\ 
-s \epsilon & c \epsilon & 0 \end{array} \right],
\end{equation}
where 
\begin{eqnarray}
\delta &=& 2 m_3 m_4 / \mu = M_Z^2 \sin 2 \beta \cos \theta_\tau / \mu, \\ 
\epsilon &=& m_5 - m_3 \mu_\tau / \mu \nonumber \\ &=& M_Z \cos \beta 
\cos \theta_\tau (\tan \theta_\tau - \mu_\tau / \mu).
\end{eqnarray}
{}From the above, $\nu_\tau$ gets a seesaw mass, i.e.
\begin{equation}
m_{\nu_\tau} = - \epsilon^2 \left( {s^2 \over M_1} + {c^2 \over M_2} \right),
\end{equation}
corresponding to the mass eigenstate
\begin{equation}
\nu'_\tau = \nu_\tau + {s \epsilon \over M_1} \tilde B - {c \epsilon \over 
M_2} \tilde W_3,
\end{equation}
whereas the other two mass eigenstates are
\begin{eqnarray}
\tilde B' &=& \tilde B + {sc \delta \over M_1 - M_2} \tilde W_3 - {s \epsilon 
\over M_1} \nu_\tau, \\ \tilde W'_3 &=& \tilde W_3 - {sc \delta \over M_1 - 
M_2} \tilde B + {c \epsilon \over M_2} \nu_\tau.
\end{eqnarray}
Because $\tilde B'$ and $\tilde W'_3$ contain $\nu_\tau$, they are not stable, 
but would decay into $\tau^\mp W^\pm$ pairs.  This may generate a lepton 
asymmetry, but it is several orders of magnitude too small because it has to 
be much less than $(\epsilon/M_{1,2})^2$, which is of order $m_{\nu_\tau}/
M_{1,2}$, i.e. $< 5 \times 10^{-13}$ if $m_{\nu_\tau} < 0.05$ eV and 
$M_{1,2} > 100$ GeV.

Consider now the couplings of $\tilde B$ and $\tilde W_3$.  The former couples 
to both $\bar \tau_L \tilde \tau_L$ and $\bar \tau^c_L \tilde \tau^c_L$, but 
the latter only to $\bar \tau_L \tilde \tau_L$, because $\tau^c_L$ is a 
singlet under $SU(2)_L$.  Since R parity is violated, both $\tilde \tau_L$ 
and $\tilde \tau^c_L$ mix with the charged Higgs boson of the supersymmetric 
standard model: $h^\pm = h_2^\pm \cos \beta + h_1^\pm \sin \beta$.  Hence 
$\tilde B'$ and $\tilde W'_3$ may decay into $\tau^\mp h^\pm$.  We now assume 
that the $\tilde \tau_L$ mixing is negligible, so that the only relevant 
coupling is that of $\tilde B$ to $\bar \tau^c_L h^+$.  Hence $\tilde W'_3$ 
decay (into $\tau^\mp h^\pm$) is suppressed because it may only do so through 
the small component of $\tilde B$ that it contains, assuming of course that 
$\tilde \tau_L$ and $\tilde \tau^c_L$ are heavier than $\tilde B$ or $\tilde 
W_3$.

We now envisage the following leptogenesis scenario.  At temperatures well 
above $T = M_{SUSY}$, the presence of sphalerons and R parity 
nonconserving interactions together ensure that there is no $L$ or $B$ 
asymmetry.  As the Universe cools down to a temperature below the masses of 
all supersymmetric particles except $\tilde B'$ and $\tilde W'_3$, we need 
only consider their interactions.  In Figure 1 we show the lepton number 
violating processes (a) $\tilde B' \leftrightarrow \tau_R^\pm h^\mp$, 
where we have adopted the more conventional notation of an outgoing $\tau_R$ 
in place of an incoming $\tau^c_L$.  These are still certainly
fast and there can be no $L$ asymmetry.  Let us assume now that $M_1 > M_2$, 
then for $T < M_1$, the $\tilde B'$ interactions are suppressed and we need 
only consider $\tilde W'_3$.  However, as we have explained in the previous 
paragraph, the lepton-number violating processes (b) $\tilde W'_3 
\leftrightarrow \tau_R^\pm h^\mp$ are slow and can satisfy the 
out-of-equilibrium condition for generating a lepton asymmetry of the 
Universe, resulting from the interference of this tree-level diagram with the 
one-loop (c) self-energy and (d) vertex diagrams.   Since 
the unsuppressed lepton number violating couplings of $\tilde B'$ are 
involved, a realistic lepton asymmetry may be generated.  It is then 
converted by the still active sphalerons into the present observed baryon 
asymmetry of the Universe.

We start with the well-known interaction of $\tilde B$ with $\tau$ and 
$\tilde \tau_R$ given by\cite{hk}
\begin{equation}
-{e \sqrt 2 \over \cos \theta_W} \left[ \bar \tau \left( {1 - \gamma_5 \over 
2} \right) \tilde B \tilde \tau_R + H.c. \right].
\end{equation}
We then allow $\tilde \tau_R$ to mix with $h^-$, and $\tilde B$ to mix with 
$\tilde W_3$, so that the interaction of the physical state $\tilde W'_3$ 
of Eq.~(16) with $\tau$ and $h^\pm$ is given by
\begin{equation}
\left( {sc\xi \over M_1 - M_2} \right) \left( {e \sqrt 2 \over \cos \theta_W} 
\right) \left[ \delta \bar \tau \left( {1 - \gamma_5 \over 2} \right) \tilde 
W'_3 h^- + H.c. \right],
\end{equation}
where $\xi$ represents the $\tilde \tau_R - h^-$ mixing and is assumed real, 
but the parameter $\delta$ of Eq.~(10) is now assumed complex.  The 
origin of this nontrivial $CP$ phase is from the $2 \times 2$ Majorana mass 
matrix spanning $\tilde B$ and $\tilde W_3$.  It is well-known that the 
off-diagonal entry is in general complex here if the diagonal entries are 
chosen to be real.

To satisfy the out-of-equilibrium condition, we require\cite{kt}
\begin{equation}
\Gamma (\tilde W'_3 \to \tau^\pm h^\mp) < H = 1.7 \sqrt {g_*} (T^2/M_{Pl})
\end{equation}
at the temperature $T \sim M_2$, where $H$ is the Hubble expansion rate of 
the Universe with $g_*$ the effective number of massless degrees of freedom 
and $M_{Pl}$ the Planck mass.  This implies
\begin{equation}
\left( {\xi |\delta| r \over M_1 - M_2} \right)^2 {(M_2^2 - m_h^2)^2 \over 
M_2^5} < 1.9 \times 10^{-14} ~{\rm GeV}^{-1},
\end{equation}
where we have used $g_* = 10^2$ and $M_{Pl} = 10^{18}$ GeV, and $r = 
(1+M_2/\mu \sin 2 \beta)/(1-M_2^2/\mu^2)$ is a correction factor for 
finite $M_2/\mu$. 
To make sure that at $T \sim M_2$, the lepton number violating processes 
$\tau^\pm h^\mp \leftrightarrow \tau^\mp h^\pm$ through $\tilde B'$ 
exchange do not erase the lepton asymmetry created by the decay of 
$\tilde W'_3$, we require
\begin{equation}
\left( {2 e^2 \xi^2 \over \cos^2 \theta_W} \right)^2 {1 \over M_1^2} 
{T^3 \over 32 \pi} {f(x) \over (1-x)^2} < H
\end{equation}
at $T \sim M_2$, where $x = M_2^2/M_1^2$ and
\begin{equation}
f(x) = 1 + {2(1-x) \over x^2} [(1+3x) \ln (1+x) - x(1+x)],
\end{equation}
which implies
\begin{equation}
{\xi^4  \over M_2} {x f(x) \over (1-x)^2} < 2.6 \times 10^{-14} 
~{\rm GeV}^{-1}.
\end{equation}
Both conditions may be satisfied for example with the following choice of 
parameters:
\begin{eqnarray}
&& \mu = 5 ~{\rm TeV}, ~~~ \sin 2 \beta = 0.5, ~~~ \xi = 2 \times 10^{-3}, 
\\ && M_1 = 3 ~{\rm TeV}, ~~~ M_2 = 2 ~{\rm TeV}, ~~~ m_h = 200 ~{\rm GeV},
\end{eqnarray}
for which the left-hand sides of Eqs.~(20) and (23) are $0.6 \times 10^{-14}$ 
GeV$^{-1}$ and $2.6 \times 10^{-14}$ GeV$^{-1}$ respectively.
 
The interference between the tree-level and self-energy + vertex diagrams 
results in the following lepton asymmetry in the decay of $\tilde W'_3$:
\begin{equation}
\epsilon = {\alpha \xi^2 \over 2 \cos^2 \theta_W} {Im \delta^2 \over 
|\delta|^2} \left( 1 - {m_h^2 \over M_2^2} \right)^2 {x^{1/2} g(x) \over 
(1-x)},
\end{equation}
where
\begin{equation}
g(x) = 1 + {2(1-x) \over x} \left[ \left( {1+x \over x} \right) \ln (1+x) 
-1 \right],
\end{equation}
which yields $3.6 \times 10^{-8}$ if $Im \delta^2 / |\delta|^2 = 1$.  
Dividing by $3 g_*$, we then obtain a realistic baryon-to-photon ratio of 
the order $10^{-10}$.

In the above, we have assumed $M_{SUSY} \sim$ few TeV.  This value is 
naturally suited for our scenario because (1) it allows $\epsilon$ 
in Eq.~(26) to be large enough without contradicting the condition given by 
Eq.~(21), and (2) it allows the lepton asymmetry to be converted at a 
temperature when the sphalerons are still very active\cite{pil,rt}, into the 
present observed baryon asymmetry of the Universe.  We have also assumed in 
our scenario that $\tilde W'_3$ only couples to $\tau^\pm h^\mp$ through its 
$\tilde B$ content.  This means that $\tilde \tau_R - h^-$ mixing is 
significant, but $\tilde \tau_L - h^-$ mixing is negligible.  The latter is 
related to the mixing of the neutrino with the neutralinos as given by 
Eq.~(9) which is indeed small and may even be set to zero, because realistic 
neutrino masses may be obtained through R parity nonconserving trilinear 
couplings instead\cite{numass}.  The former is usually taken\cite{mix} from 
the soft supersymmetry breaking term $h_1^- \tilde \nu_\tau \tilde \tau^c_L$ 
and the $\tilde \tau_L h_2^+$ term through $\tilde \tau_L^* - \tilde 
\tau_L^c$ mixing.  In that case, the $\tilde \tau_R - h^-$ mixing would be 
proportional to a linear combination of $\langle \tilde \nu_\tau \rangle$ 
and $\mu_\tau$, and be constrained to be very small\cite{future}. 
For our scenario to be successful, we need to introduce the new term
\begin{equation}
H_2^\dagger H_1 \tilde \tau^c_L = (\bar h_2^+ h_1^0 + \bar h_2^0 h_1^-) 
\tilde \tau^c_L,
\end{equation}
which is unconstrained and may have the desired mixing of $\xi = 2 \times 
10^{-3}$ shown.  Such a soft supersymmetry breaking term is 
nonholomorphic\cite{hr} and whereas this class of couplings has been studied 
previously\cite{bfpt}, we are the first to make use of this particular term.  
We should also mention that although we have used only 
$\tau$ to represent a lepton, replacing it with $e$ or $\mu$ is just as 
appropriate.

In conclusion, we have shown in this paper how R parity nonconservation may 
create its own successful leptogenesis through the suppressed decay of 
$\tilde W'_3$ into $l^\pm h^\mp$.  We require $M_{SUSY} \sim$ few TeV and 
the existence of the nonholomorphic soft supersymmetry breaking term 
$H_2^\dagger H_1 \tilde l^c_L$.  Whereas this mechanism depends on R parity 
nonconservation, it comes from a sector distinct from that responsible for 
nonzero neutrino masses\cite{bb}.  We also require the soft supersymmetry 
breaking gaugino masses $M_1$ and $M_2$ to be a few TeV with $M_1 > M_2$.

E.M. thanks S. Pakvasa and X. Tata for discussions, and the support of 
the U.S.~Department of Energy under Grant No. DE-FG03-94ER40837.

\bibliographystyle{unsrt}

\begin{thebibliography}{99}

\bibitem{ma}  
 For a recent overview, see for example E. Ma, Phys. Rev. Lett. {\bf 81}, 
  1171 (1998).

\bibitem{seesaw} 
 M. Gell-Mann, P. Ramond and R. Slansky, 
  in {\it Supergravity}, edited by P. van Nieuwenhuizen and D. Freedman, 
  (North-Holland, 1979), p.~315;
 T. Yanagida, in {\it Proceedings of the Workshop on the Unified Theory 
  and the Baryon Number in the Universe}, edited by O. Sawada and 
  A. Sugamoto (KEK Report No.~79-18, Tsukuba, 1979), p.~95;
 R. N. Mohapatra and G. Senjanovi\'{c}, Phys. Rev. Lett. {\bf 44},
  912 (1980). 

\bibitem{triplet} 
 E. Ma and U. Sarkar, Phys. Rev. Lett. {\bf 80}, 5716 (1998).

\bibitem{fy} 
 M. Fukugita and T. Yanagida, Phys. Lett. {\bf 174B}, 45 (1986).

\bibitem{krs}
  V. A. Kuzmin, V. A. Rubakov and M. E. Shaposhnikov, 
   Phys. Lett. {\bf 155B}, 36 (1985).

\bibitem{bary}
  M. Fukugita and T. Yanagida, Phys. Rev. {\bf D42}, 1285 (1990); 
  J. A. Harvey and M. S. Turner, Phys. Rev. {\bf D42}, 3344 (1990); 
  A. E. Nelson and S. M. Barr, Phys. Lett. {\bf B246}, 141 (1990); 
  W. Buchm\"{u}ller and T. Yanagida, Phys. Lett. {\bf B302}, 240 (1993).

\bibitem{numass}
  R. Hempfling, Nucl. Phys. {\bf B478}, 3 (1996);
  F. M. Borzumati, Y. Grossman, E. Nardi, and Y. Nir, Phys. Lett. {\bf B384}, 
   123 (1996);
  E. Nardi, Phys. Rev. {\bf D55}, 5772 (1997);
  M. A. Diaz, J. C. Romao, and J. W. F. Valle, Nucl. Phys. {\bf B524}, 23 
   (1998);
  M. Drees, S. Pakvasa, X. Tata, and T. ter Veldhuis, Phys. Rev. {\bf D57}, 
   R5335 (1998);
  B. Mukhopadhyaya, S. Roy, and F. Vissani, Phys. Lett. {\bf B443}, 191 (1998);
  E. J. Chun, S. K. Kang, and C. W. Kim, Nucl. Phys. {\bf B544}, 89 (1999); 
  R. Adhikari and G. Omanovic, Phys. Rev. {\bf D59}, 073003 (1999); 
  O. C. W. Kong, Mod. Phys. Lett. {\bf A14}, 903 (1999);
  S. Rakshit, G. Bhattacharyya, and A. Raychaudhuri, Phys. Rev. {\bf D59}, 
   091701 (1999);
  S. Y. Choi, E. J. Chun, S. K. Kang, and J. S. Lee, Phys. Rev. {\bf D60}, 
   075002 (1999);
  L. Clavelli and P. H. Frampton, hep-ph/9811326;
  D. E. Kaplan and A. E. Nelson, hep-ph/9901254;
  A. S. Joshipura and S. Vempati, hep-ph/9903435;
  Y. Grossman and H. E. Haber, hep-ph/9906310;
  G. Bhattacharyya, H. V. Klapdor-Kleingrothaus, and H. Pas, Phys. Lett. 
   {\bf B463}, 77 (1999);
  A. Abada and M. Losada, hep-ph/9908352;
  O. Haug, J. D. Vergados, A. Faessler, and S. Kovalenko, hep-ph/9909318;
  E. J. Chun and S. K. Kang, hep-ph/9909429;
  F. Takayama and M. Yamaguchi, hep-ph/9910320.

\bibitem{erase}
  B. A. Campbell, S. Davidson, J. E. Ellis, and K. Olive, Phys. Lett. {\bf 
   B256}, 457 (1991);
  W. Fischler, G. F. Giudice, R. G. Leigh, and S. Paban, Phys. Lett. {\bf 
   B258}, 45 (1991);
  E. Ma, M. Raidal, and U. Sarkar, Phys. Lett. {\bf B460}, 359 (1999).

\bibitem{maroy}
  E. Ma and P. Roy, Phys. Rev. {\bf D41}, 988 (1990).

\bibitem{cf}
  C.-H. Chang and T.-F. Feng, hep-ph/9901260; see also 
  M. A. Diaz, J. C. Romao, and J. W. F. Valle, Ref.\cite{numass}.

\bibitem{hk} 
  H. E. Haber and G. L. Kane, Phys. Rep. {\bf 117}, 75 (1985).

\bibitem{kt}
  E. W. Kolb and M. S. Turner, {\em The Early Universe} (Addison-Wesley, 
   Reading, MA, 1990).

\bibitem{pil}
  See for example A. Pilaftsis, Phys. Rev. {\bf D56}, 543 (1997); Int. J. 
   Mod. Phys. {\bf A14}, 1811 (1999).

\bibitem{rt}
  Note also that we do not require the electroweak phase transition to be 
   first-order for satisfying the out-of-equilibrium condition.  See for 
   example A. Riotto, hep-ph/9807454; A. Riotto and M. Trodden, 
   hep-ph/9901362.

\bibitem{mix}
  A. Akeroyd, M. A. Diaz, J. Ferrandis, M. A. Garcia-Jareno, and J. W. F. 
   Valle, Nucl. Phys. {\bf B529}, 3 (1998); 
  C.-H. Chang and T.-F. Feng, hep-ph/9908295.  

\bibitem{future}
  This and other details will be discussed elsewhere.

\bibitem{hr}
  L. J. Hall and L. Randall, Phys. Rev. Lett. {\bf 65}, 2939 (1990).

\bibitem{bfpt}
  F. Borzumati, G. R. Farrar, N. Polonsky, and S. Thomas, Nucl. Phys. 
   {\bf B555}, 53 (1999).

\bibitem{bb}
  This is not unlike the case of neutrinoless double beta decay, where the 
dominant contribution may not be directly related to nonzero Majorana neutrino 
masses.

\end{thebibliography}

\begin{figure}[htb]
\mbox{}
\vskip 3.6in\relax\noindent\hskip .5in\relax
\includegraphics{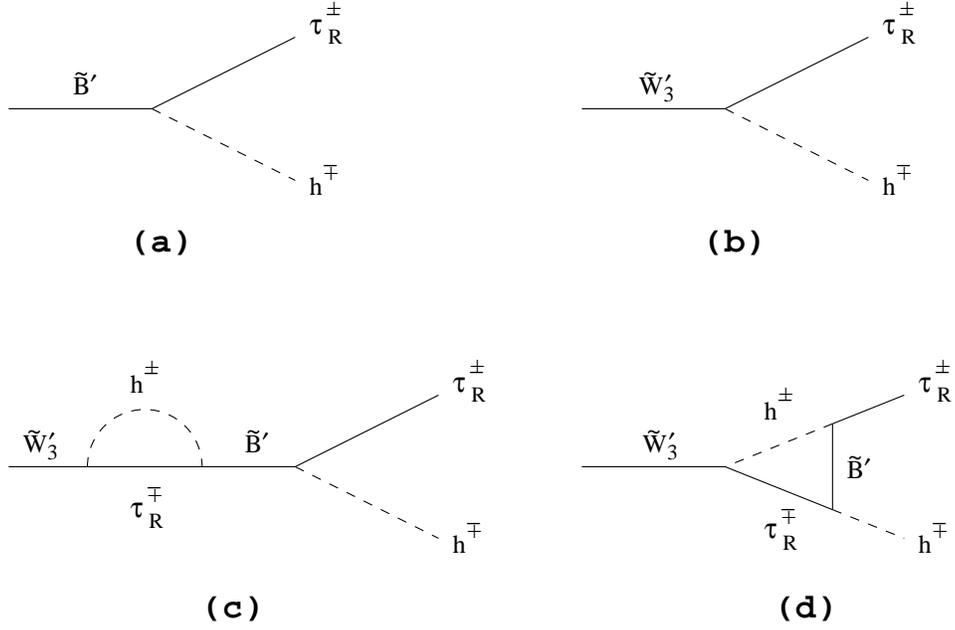} 
\vskip .25in
\caption{ Tree-level diagrams for (a) $\tilde B'$ 
decay and (b) $\tilde W'_3$ decay (through their $\tilde B$ content), 
and the one-loop (c) self-energy and (d) vertex diagrams for  
$\tilde W'_3$ decay which have absorptive parts of opposite lepton number.}
\end{figure}

\end{document}